\documentclass[fleqn,aps,pra,amsmath,amssymb,reprint,showpacs]{revtex4-1}
\usepackage{amsmath}
\usepackage{graphicx}
\usepackage{dcolumn}
\usepackage{bm}
\usepackage{textcomp}
\usepackage{natbib}
\usepackage{enumerate}

\def\ket#1{\mathinner{|{#1}\rangle}}

\renewcommand\Im{\operatorname{Im}}
\begin{document}
\title{Coherent population trapping (CPT) versus electromagnetically induced transparency (EIT)} 
\author{Sumanta Khan}
\author{Molahalli Panidhara Kumar}
\author{Vineet Bharti}
\author{Vasant Natarajan}
\affiliation{Department of Physics, Indian Institute of Science, Bangalore 560012, India}

\begin{abstract}
We discuss the differences between two well-studied and related phenomena---coherent population trapping (CPT) and electromagnetically induced transparency (EIT). Many differences between the two---such as the effect of power in the beams, detuning of the beams from resonance, and the use of vapor cells filled with buffer gas---are demonstrated experimentally. The experiments are done using magnetic sublevels of the $ 1 \rightarrow 1 $ transition in the D$_2$ line of $^{87}$Rb. \\

\end{abstract}

\maketitle

\section{Introduction}
The phenomenon of coherent population trapping (CPT), reviewed by Arimondo in Ref.\ \cite{ARI96}, and the phenomenon of electromagnetically induced transparency (EIT), reviewed in Refs.\ \cite{HAR97} and \cite{FIM05}, are two sides of the same coin. Both phenomena have been studied for a long time in three-level systems---CPT in lambda type; and EIT in all the three kinds namely lambda type, ladder type, and vee type. The physics underlying the two phenomena are related, but are also distinct in many ways. The aim of this article is to highlight the differences between the two, partly to eliminate the confusion existing in the literature. Some of this confusion arises due to the use of the term EIT for what are really CPT processes.

CPT was first observed as a decrease in the fluorescence emission from Na atoms in a vapor cell \cite{AGM76}. It was understood to arise from the fact that the atoms were being optically pumped into a dark non-absorbing state by the excitation beams. Once pumped, atoms were trapped in that state because there was no way to spontaneously decay from there---hence the name {\bf coherent population trapping}. 

\begin{figure}
\centering{\resizebox{0.75\columnwidth}{!}{\includegraphics{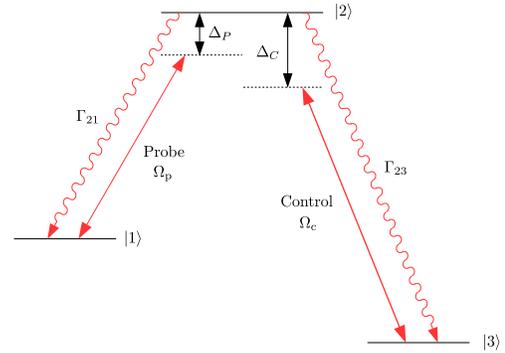}}}
\caption{(Color online) Three-level $\Lambda$-type system. The different parameters are explained in the text.}
 \label{lambda}
\end{figure}

The dark state was created by destructive interference between excitation pathways from two stable levels to a common excited level. The three levels form the required $\Lambda$-type system, as shown in Fig.\ \ref{lambda}. It is customary to call the laser driving the $\ket{1} \leftrightarrow \ket{2}$ transition as the \textbf{probe} laser, and the one driving the $\ket{2} \leftrightarrow \ket{3}$ transition as the \textbf{control} laser. Both transitions are electric-dipole (E1) allowed transitions; and thus the transition $\ket{1} \leftrightarrow \ket{3}$ is E1 forbidden. The detunings are labeled by $\Delta$'s, the powers by the respective Rabi frequencies $\Omega$'s, and the spontaneous decay rates by $\Gamma$'s. $\Gamma_{21}/2 \pi$ and $\Gamma_{23}/2 \pi$ (i.e.\ for the E1 allowed transitions) are about 1--10 MHz, while $\Gamma_{31}$ for the forbidden transition is close to 0. In reality, there is decoherence between the levels due to inter-atomic collisions, collisions with walls, spontaneous scattering through the upper level, etc. It is important to note that the two lasers are required to be {\bf phase coherent} so that the dark superposition state can be formed.

EIT works on a slightly different principle. The phenomenon is usually studied in the regime where the control laser is {\bf strong} and the probe laser is {\bf weak}, i.e.\ $\Omega_p \ll \Omega_c$. Under these conditions, the control laser shifts the energy levels of the atom away from line center through the AC Stark effect, which can also be understood as the creation of new {\bf dressed states} of the coupled atom-photon system \cite{COR77}. The shift is equal to the Rabi frequency of the control laser. The absorption of the probe laser splits into a classic Autler-Townes doublet with a splitting equal to the control Rabi frequency, and shows enhanced transparency at line center---a transparency induced by the control laser. Hence the name {\bf electromagnetically induced transparency}. Quantum interference also plays a role in EIT, but mainly when the control Rabi frequency is of the order of the excited-state linewidth. The resulting (destructive) interference between the transition amplitudes from the ground state to the control-induced dressed states is particularly important in lambda systems, as discussed in Ref.\ \cite{LIX95}. The interference then leads to the probe absorption being identically zero at line center. Dressed-state interference is not so important in vee and ladder systems \cite{KBN16}. In addition, since the probe photon is not involved in the interference, the probe beam in EIT experiments can be (and is) generally derived from a {\bf phase independent} laser.

The above analysis shows that both EIT and CPT are two-photon processes, because the transitions involve two different levels. However, in CPT both the probe and control photons are equally important. By contrast, in EIT the transparency is entirely created by the strong control laser, while the weak probe laser only plays the role of measuring this transparency. With these basic differences in mind, let us now contrast the two phenomena in terms of various experimental parameters.
\begin{enumerate}

\item {\bf Scan axis.} An important difference between EIT and CPT is the scan axis. In CPT, this is the relative detuning between the two beams. 
If the experiment is done using magnetic sublevels of a degenerate transition, then the frequency difference between the two beams is only of the order of few MHz. The required phase coherence can then be achieved using an acoustic-optic modulator (AOM), and the scan axis is the frequency of the AOM driver. 

The CPT phenomenon can also be used to access the clock transition in alkali atoms like Rb and Cs. In this case, the probe and control beams differ in frequency by the ground hyperfine interval, which is of order few GHz and hence too large to be achieved with an AOM. Therefore, it is done in one of the two following ways:
\begin{enumerate}[(i)]
	\item Taking two independent lasers and beating their outputs on a sufficiently fast photodiode, as has been done in Ref.\ \cite{BNW97}. In this case, the scan axis is the frequency of the beat signal.
	
	\item Taking one laser and feeding its output through an electro-optic modulator (EOM). The EOM driver is set to half the clock frequency because it produces two sidebands spaced apart by twice the EOM frequency. The scan axis is the frequency of the EOM driver.
\end{enumerate}

On the other hand, the scan axis in EIT is the frequency of the phase-independent probe laser. If the transition is in the optical regime, the frequency is of order $10^{15}$ Hz, or 6 to 9 orders-of-magnitude larger than that for CPT.

\item {\bf Subnatural resonance---the relevant natural \\ linewidth $\Gamma$.} The discussion of the scan axis brings us to the question of what defines whether the feature is subnatural or not. In CPT experiments, the scanning device is at the difference frequency between the two lower levels. The relevant natural linewidth is therefore the linewidth of transition between these two levels, which as mentioned before is close to zero because the transition is electric-dipole forbidden. The linewidth of the upper level does not enter the picture. As an example, we consider the observation of a resonance linewidth of 42 Hz for coherent dark resonances in a Cs vapor cell filled with buffer gas \cite{BNW97}. The resonance is on the ground hyperfine splitting of $^{133}$Cs---the clock transition at 9.1 GHz used in the SI definition of the second. The experiments were done on the $\rm D_2$ line of Cs, but the 5~MHz linewidth of the upper state does not enter the picture. Nowhere in the above work do the authors call their resonance {\bf subnatural}, because it is well known that the natural linewidth of the clock transition is well below 1 Hz. Similarly, the phase-locked lasers used in the above study have a linewidth of order 1 MHz. This does not prevent them from seeing a 42 Hz feature because they are looking for a beat signal between two phase-coherent lasers, which can be much narrower than the linewidth of the laser. This shows that the CPT phenomenon can be used for precision spectroscopy on the ground levels \cite{WYN99}.

On the other hand, we have shown that the narrow features of EIT are really {\bf subnatural} and can be used for high-resolution spectroscopy of the upper level \cite{RAN02}. This is because the scan axis is the optical frequency of the probe laser, and the relevant natural linewidth is the linewidth of the upper level. In the case of the Rb $\rm D_2$ line used in that work, the upper level had a linewidth of 6~MHz. Therefore, any feature that is narrower than 6 MHz would be called subnatural. The narrowest feature that we have reported is 0.85 MHz or $\Gamma/7$ \cite{IKN08}. This corresponds to a $Q$ of $4.5 \times 10^8$, which is better than the $Q$ of $2.2 \times 10^8$ for the CPT experiment on the Cs clock transition in Ref.\ \cite{BNW97}.

\item {\bf Fluorescence versus absorption.} One important consequence of the fact that CPT results from the creation of a dark state is that there is a concomitant decrease in the fluorescence from the cell---it becomes \textbf{dark}. Recall that the first observation of CPT was the appearance of a dark region in the fluorescence from a Na vapor cell \cite{AGM76}. In that study, an inhomogeneous magnetic field was applied along the axis of the cell, so that the dark state was created only in a small region. As a result, the CPT phenomenon was seen as a dark line in a bright cell. By contrast, in EIT, the strong control laser is always being absorbed (and atoms are fluorescing), and the induced transparency is seen only in the absorption signal of the weak probe laser. Therefore, at least to first order, there is no change in the fluorescence whether the probe laser is on or off.

\item {\bf Effect of buffer gas.} Because CPT is a ground-state coherence phenomenon, any technique that increases the coherence time will give a narrower linewidth. One of the most common methods is to use a buffer gas in the vapor cell---typically a few torrs of a gas like Ne or N$_2$. For example, the CPT experiment on the clock transition in Ref.\ \cite{BNW97} was done in a Cs vapor cell filled with N$_2$ buffer gas. On the other hand, the use of such cells for EIT experiments actually kills the signal, because the buffer gas considerably broadens the normal probe absorption signal due to collisions and swamps any modification due to the control.

\item {\bf Power in the two beams}. As mentioned before, both beams play an equally important role in CPT, while EIT is studied in the regime where the control is strong and the probe is weak. This leads to an important difference---there is not much point in increasing the control power beyond the probe power in CPT. The role of control power is to increase the linewidth due to increased decoherence through the upper level---\textbf{power broadening}---which increases the spontaneous scattering rate, and the narrowest feature is obtained when the two powers are equal.

\item {\bf Effect of detuning.} The upper level in CPT causes decoherence (and hence broadens the resonance) by inducing real transitions from the lower levels. This transition rate (and hence linewidth) can be reduced by detuning from the level. But the reduced transition probability also results in a smaller signal strength and signal-to-noise ratio (SNR). Note that the CPT resonance still occurs at the same {\bf relative} detuning between the two beams, which is the point at which the two-photon Raman resonance condition is satisfied. On the other hand, detuning the control laser from the upper level in EIT causes the resonance to shift within the absorption profile of the probe laser. The resonance again appears where the probe detuning matches the control detuning, but the detuned {\bf lineshape} is very different.

\end{enumerate}
Many of the above points will be validated below through experiments on the $\rm D_2$ line of $^{87}$Rb.

\section{Experimental details}

\begin{figure}
\centering{\resizebox{0.75\columnwidth}{!}{\includegraphics{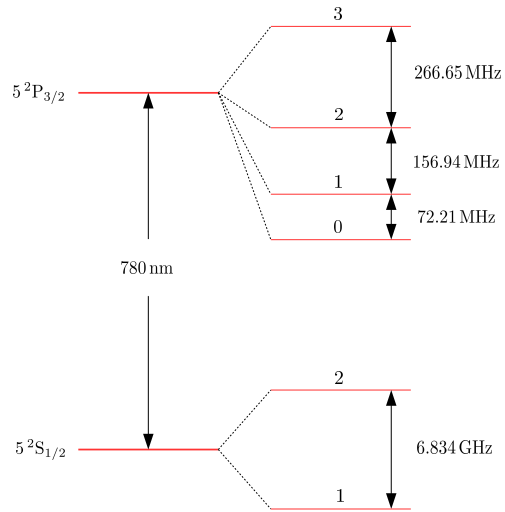}}}
\caption{(Color online) Low-lying energy levels of the D$_2$ line of $^{87}\rm Rb $ used in the experiment. The $F$ value of each level is indicated.}
 \label{rubidium_87_levels}
\end{figure}

The relevant energy levels of the $^{87}$Rb D$_2$ line ($ 5\rm{S}_{1/2} \rightarrow 5\rm{P}_{3/2} $ transition) at 780 nm used for the EIT and CPT experiments are shown in Fig.\ \ref{rubidium_87_levels}. 

\begin{figure}
\centering{\resizebox{0.95\columnwidth}{!}{\includegraphics{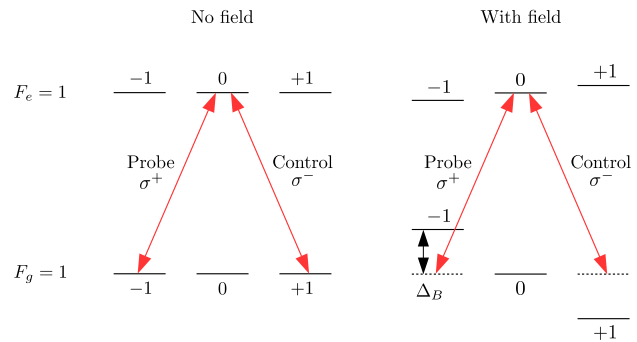}}}
\caption{(Color online) Magnetic sublevels of the $ 1 \rightarrow 1 $ transition used in the experiment. The beams get detuned in the presence of a magnetic field, as shown on the right.}
 \label{magneticlevels}
\end{figure}

The required $\rm \Lambda $-type system is formed using magnetic sublevels of the $ 1 \rightarrow 1 $ transition along with opposite circular polarizations for the control and probe beams. The natural linewidth of the upper state is 6 MHz, and the saturation intensity---the intensity at which the transition gets power broadened by a factor of $\sqrt{2}$ ---is given by (page 209 of Ref.\ \cite{NAT15})
\begin{equation*}
I_s = \dfrac{\pi h c}{3 \lambda^3 \tau}
\end{equation*}
where $\lambda =  780$ nm is the wavelength of the transition and $\tau = 26.25$ ns is the lifetime of the excited state \cite{GAF02}. The value for this level is 1.67 mW/cm$^2$. The resultant scheme is shown in Fig.\ \ref{magneticlevels}. The figure also shows that applying a magnetic field is equivalent to detuning the beams from resonance. Thus, the effect of detuning can be studied by applying a magnetic field. 

\begin{figure}
(a)\centering{\resizebox{0.9\columnwidth}{!}{\includegraphics{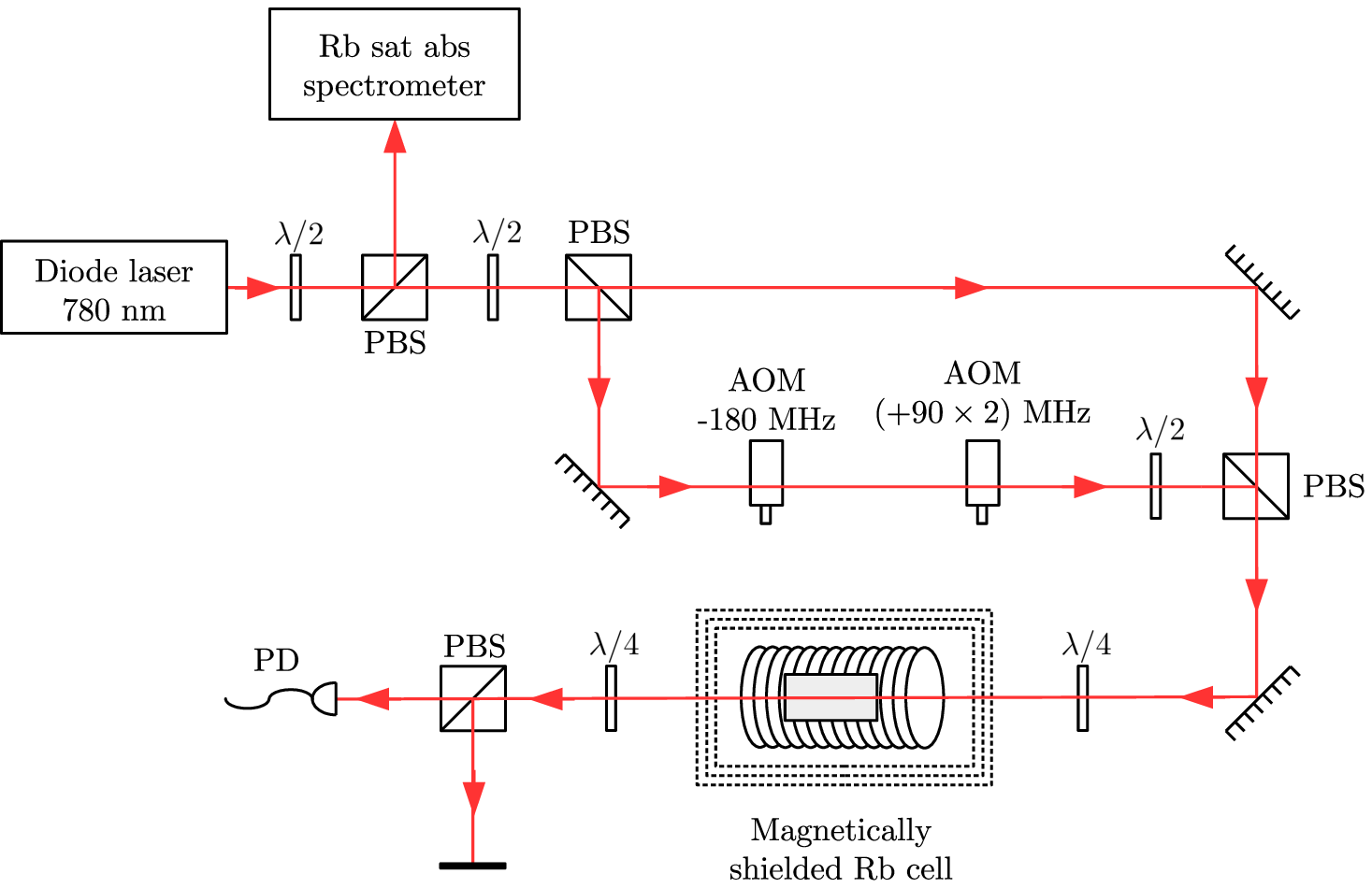}}}
(b)\centering{\resizebox{0.9\columnwidth}{!}{\includegraphics{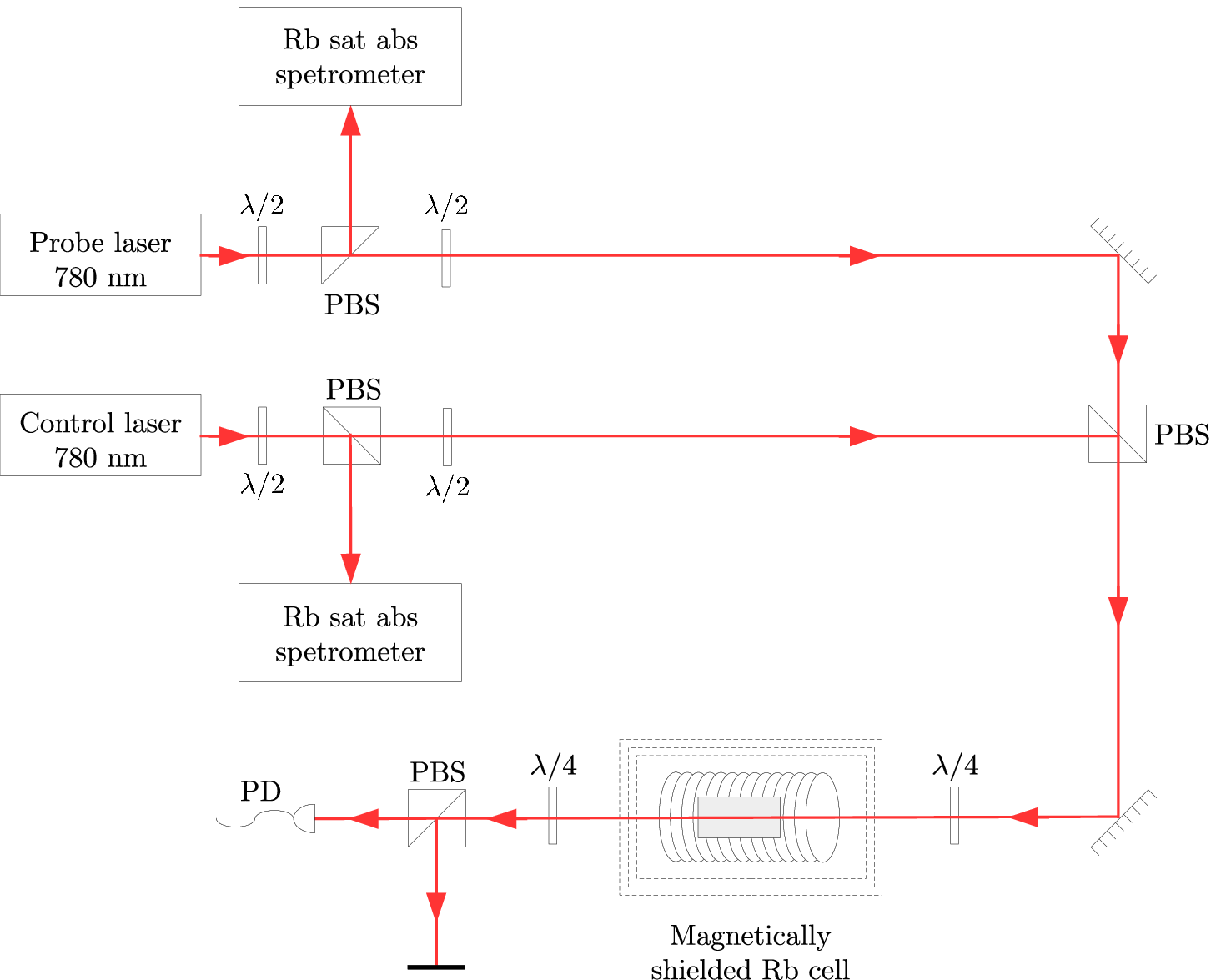}}}
\caption{(Color online) Experimental schematic for CPT and EIT experiments. Figure key: $ \lambda/2 $ -- half wave retardation plate; PBS -- polarizing beam splitter cube; $\lambda/4$  -- quarter wave retardation plate; AOM -- acousto-optic modulator; PD -- photodiode. (a) The required phase coherence for CPT experiments is achieved by deriving both beams from a single laser. The probe beam is locked while the control beam is scanned by scanning the frequency of the double-passed AOM. (b) EIT experiments with independent control and probe lasers. In this case, the probe beam is scanned while the control beam is locked.}
 \label{schema}
\end{figure}

The experimental setup for the CPT experiment is shown schematically in Fig.\ \ref{schema}(a). The required phase coherence for the control and probe beams is obtained by deriving both beams from a single laser, and using AOMs to produce the required frequency shift. The laser consists of a grating stabilized diode laser system, as described in Ref.\ \cite{MRS15}. The linewidth of the laser after stabilization is 1 MHz. The output beam is elliptic, and has a size of 3 mm $\times$ 4 mm. Since magnetic sublevels are used, the two beams are required to be at the same frequency when no field is present. This is achieved by using two AOMs in the path of the control beam---one AOM with a downshift of 180 MHz, and the other compensating for this shift by having a double-passed upshift of 90 MHz. The double passing ensures that the direction of the beam does not change when it is scanned. The frequency of the AOMs is set using function generators.

The power in the beams is controlled using $ \lambda /2 $ waveplates in front of respective PBSs. The two beams have orthogonal linear polarizations, and are mixed using a PBS. Their polarizations are changed to circular using a $ \lambda/4 $ waveplate in front of the vapor cell. The vapor cell is cylindrical with dimensions of 25 mm diameter $ \times $ 50 mm length. The cell is inside a three-layer magnetic shield that reduces stray fields to $< 1$ mG. Two kinds of cells are used---one that is pure and contains both isotopes in the natural abundances, and the second that contains only $^{87}$Rb with 20 torr of Ne as buffer gas. 

The laser is locked to the $ 1 \rightarrow 1 $ transition using the saturated absorption spectroscopy (SAS) signal from another vapor cell. This ensures that the (unshifted) probe beam is locked to the $ m_F = -1 \rightarrow m_F = 0 $ transition. The control beam is scanned around the $ m_F = +1 \rightarrow m_F = 0 $ transition by scanning the frequency of the double-passed AOM. The polarizations after the cell are made linear using a second $ \lambda/4 $ waveplate, and the beams are separated using another PBS. The probe beam alone is detected using a photodiode---therefore, the photodiode signal is proportional to probe transmission. Since the SAS signal used for locking corresponds to absorption by zero-velocity atoms, detecting the non-scanning probe beam allows us to have a flat Doppler-free background for the CPT signal.

The schematic for the EIT experiments is shown in Fig.\ \ref{schema}(b). This is essentially the same as that used in the CPT experiments, except that the two beams are derived from independent laser systems. In this case, the control laser is locked to the $1 \rightarrow 1 $ transition using an SAS signal, which ensures that the control beam is resonant with the $m_F =+1 \rightarrow m_F = 0 $ transition. The probe laser is scanned around the same  $ 1 \rightarrow 1 $ transition by varying the angle of the grating, which means that the EIT spectrum appears on a Doppler pedestal.

The absorption through a normal vapor cell is about 5\%. Therefore, the photodiode signal is multiplied suitably to get this degree of absorption off resonance. The absorption through a buffer-gas filled cell is about a factor of 2 lower. Hence the photodiode signal is multiplied by a suitably smaller number.

\section{Results and discussion}

\subsection{Saturated absorption spectroscopy (SAS)}

As mentioned above, laser locking to the relevant hyperfine transition was done using an SAS signal. The SAS spectrometer consists of a low-power probe beam ($\sim $ 20 \textmu W), whose absorption is saturated by a strong pump beam ($\sim$ 2 mW) which is counter-propagating with respect to the probe. Probe transmission then shows Doppler-free peaks at the location of various hyperfine transitions (page 309 of Ref.\ \cite{NAT15}). This is because only zero-velocity atoms are resonant with both beams. 

\begin{figure}
\centering{\resizebox{0.95\columnwidth}{!}{\includegraphics{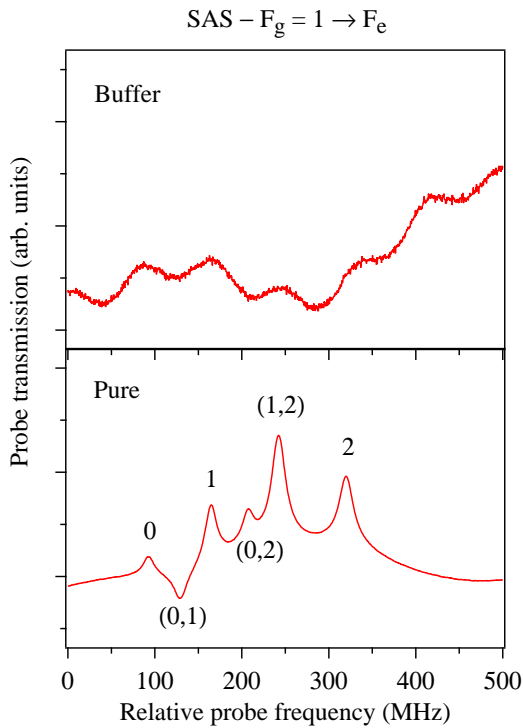}}}
\caption{(Color online) Doppler subtracted SAS signals for $ 1 \rightarrow F_e $ transitions obtained in the two kinds of cells.}
 \label{sas}
\end{figure}

In order to make the spectrum appear on a flat background, the signal from a second identical probe beam is subtracted from the main one. The resulting spectra for both kinds of cells are shown in Fig.\ \ref{sas}. The spectrum obtained in a pure cell shows six peaks---three corresponding to real hyperfine transitions, and three spurious ``crossover'' resonances. The crossover resonances appear exactly halfway between two real peaks, and occur because of interaction with non-zero-velocity atoms in the vapor cell.

The spectrum obtained in a buffer cell, taken under identical conditions, shows a complete lack of any peaks. This shows unequivocally that such a cell cannot be used for EIT experiments. However, as we will see later, they are quite useful for CPT experiments, because collisions with a buffer gas increase the coherence time and result in a significant reduction of the linewidth.

\subsection{CPT and EIT in a pure cell}

\begin{figure}
\centering{\resizebox{0.95\columnwidth}{!}{\includegraphics{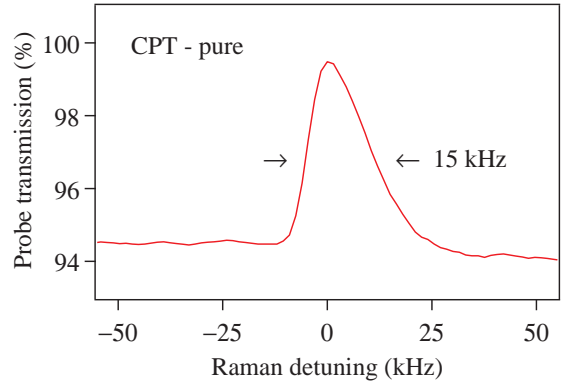}}}
\caption{(Color online) CPT resonance obtained in a pure cell. The background level is flat.}
 \label{cptpure}
\end{figure}

The CPT resonance obtained in a pure cell as a function of the relative (Raman) detuning between the beams is shown in Fig.\ \ref{cptpure}. The probe transmission signal shows a peak when the detuning is zero---the point at which the dark state is formed. The linewidth of the resonance is 15 kHz, which is not \textbf{subnatural} because the natural linewidth between the lower levels is close to 0. The observed linewidth is actually determined by non-radiative decoherence processes---such as background collisions, spontaneous scattering through the upper level, and transit time across the beams. The 1 MHz linewidth of the laser does not prevent the observation of such a narrow linewidth because the path length difference between the probe and control beams when they reach the atoms is negligible; certainly orders-of-magnitude less than the coherence length of the laser. The signal level away from resonance (background level) is flat, which is because the probe beam is locked and hence only addresses zero-velocity atoms.

\begin{figure}
\centering{\resizebox{0.95\columnwidth}{!}{\includegraphics{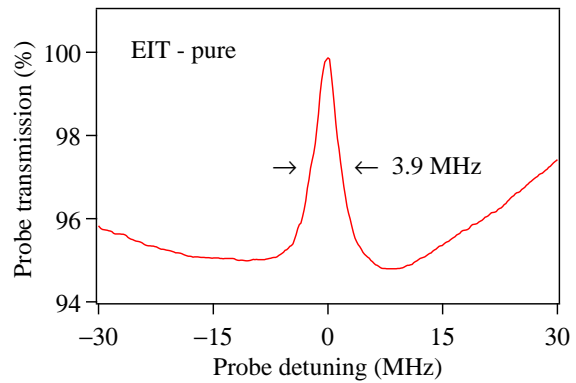}}}
\caption{(Color online) EIT resonance obtained in a pure cell. The background level is curved because of the Doppler effect.}
 \label{eitpure}
\end{figure}

The EIT resonance as a function of probe detuning is shown in Fig.\ \ref{eitpure}. The resonance also appears as a peak in the probe transmission spectrum, but there are important differences from the CPT case. First, the linewidth is 3.9 MHz, which is \textbf{subnatural} given the 6 MHz linewidth of the excited state. Secondly, the background level away from resonance has a Doppler profile because the probe beam is being scanned.

\subsection{Effect of control power}

In CPT experiments, the control beam causes decoherence through the upper level by increasing the scattering rate. The scattering rate is given by 
\begin{equation}
R = \dfrac{\Gamma}{2} \dfrac{I/I_s}{1 + I/I_s} 
\label{scrate}
\end{equation}
The above equation shows that the scattering rate increases initially and then reaches a saturation value as the intensity is increased. Therefore, the linewidth will show a similar behavior and reach a saturation value when the control intensity is high.

\begin{figure}
\centering{\resizebox{0.85\columnwidth}{!}{\includegraphics{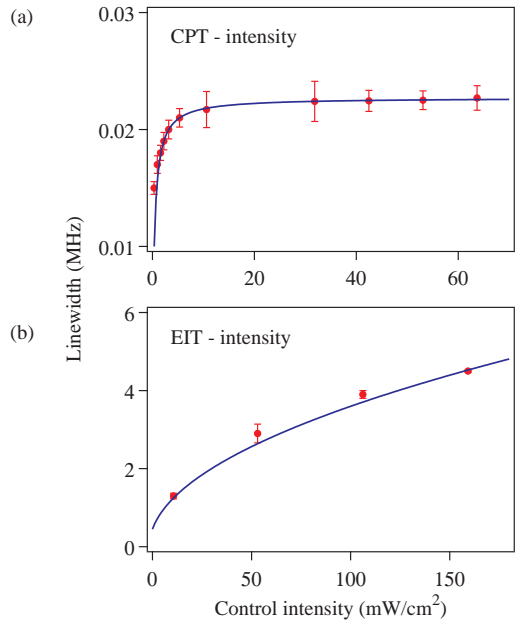}}}
\caption{(Color online) Effect of control intensity on the linewidth of the resonance. The power is converted to an average intensity by dividing the maximum intensity at the center of the beam by 2. (a) For CPT resonances, showing increase in linewidth due to increased decoherence through the upper level. The solid line is a fit to the scattering-rate expression in Eq.\ \eqref{scrate}. (b) For EIT resonances, showing increase in linewidth due to power broadening. The solid line is a fit to the power-broadening expression in Eq.\ \eqref{power}.}
 \label{powervar}
\end{figure}

The results are shown in Fig.\ \ref{powervar}(a). The solid line is the fit to Eq.\ \eqref{scrate}, and shows that the measured data follows the expected behavior. However, the saturation intensity obtained from the fit is about 0.5 mW/cm$^2$, which neither corresponds to that of the upper state nor to that determined by the decoherence rate between the lower levels, but is some combination of the two.

On the other hand, the strong control beam in EIT experiments causes power broadening of the upper level. For a control intensity of $ I $, the power-broadened linewidth is given by
\begin{equation}
\Gamma = \Gamma_{\circ} \sqrt{1 + \dfrac{I}{I_s}}
\label{power}
\end{equation}
where $ \Gamma_{\circ} $ is the natural linewidth and $I_s $ is the saturation intensity for the transition. Thus, as the control power increased, the linewidth of the upper level, and consequently the linewidth of the EIT resonance, increases. These results are shown in Fig.\ \ref{powervar}(b). The solid line is a fit to the above equation, which yields a value of $ I_s = 1.59(2)$ mW/cm$^2$. This value is close to the 1.67 mW/cm$^2$ saturation intensity of the upper state, which is as expected because the dressed states created by the control laser involve the upper state.

\subsection{Effect of detuning}
As seen from the magnetic sublevel structure shown in Fig.\ \ref{magneticlevels}, the two beams get detuned in the presence of a magnetic field. 

\begin{figure}
\centering{\resizebox{0.95\columnwidth}{!}{\includegraphics{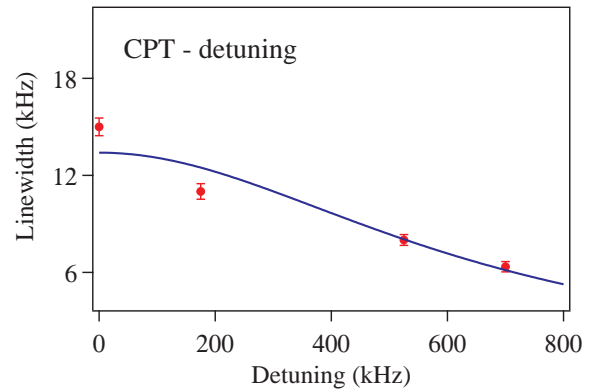}}}
\caption{(Color online) Effect of detuning on CPT resonances. The linewidth decreases because of reduced scattering from the upper level. The solid line is a fit to Eq.\ \eqref{scratedelta}, which gives the scattering rate in the presence of detuning.}
 \label{cpt_detuning}
\end{figure}

For CPT, the scattering rate given by Eq.\ \eqref{scrate} in the presence of detuning $\delta$ changes to 
\begin{equation}
R_{\delta}= \dfrac{\Gamma}{2} \dfrac{I/I_s}{1 + I/I_s + 4 \delta^2/\Gamma^2} 
\label{scratedelta}
\end{equation}
This shows that the scattering rate decreases as the detuning is increased---hence, the linewidth of the CPT resonance will also decrease. The results are shown in Fig.\ \ref{cpt_detuning}. The solid line, which is a fit to the above equation, shows that the reduced scattering rate describes the data well.

\begin{figure}
\centering{\resizebox{0.95\columnwidth}{!}{\includegraphics{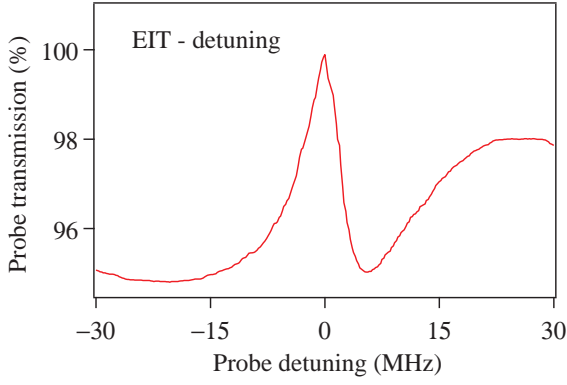}}}
\caption{(Color online) EIT resonance in the presence of detuning, achieved with a magnetic field of 1 G corresponding to a level shift of 0.7 MHz. The lineshape of the resonance is highly asymmetric.}
 \label{eit_detuning}
\end{figure}

On the other hand, the lineshape of the detuned EIT resonance becomes asymmetric and non-Lorentzian. This is because the detuned EIT resonance moves inside the absorption profile. The lineshape of the resonance, shown in Fig.\ \ref{eit_detuning}, confirm this explanation. It is incorrect to define a linewidth for the asymmetric lineshape, and hence it is not done.

\subsection{Effect of buffer gas}

\begin{figure}
\centering{\resizebox{0.95\columnwidth}{!}{\includegraphics{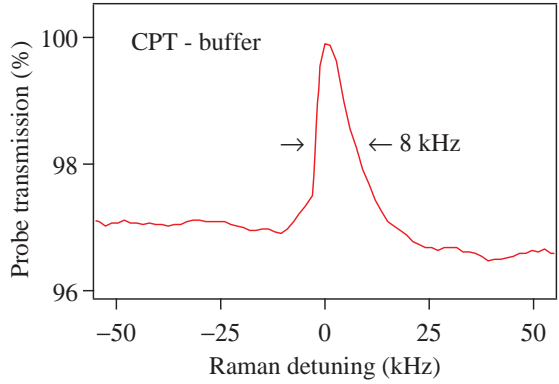}}}
\caption{(Color online) CPT resonance in a buffer-gas filled cell. The features are the same as that obtained in a pure cell, except that the linewidth is reduced by a factor of 2.}
 \label{cptbuffer}
\end{figure}

The presence of a buffer gas in the vapor cell results in an increase in the coherence time of the lower levels. This is advantageous for CPT experiments because it results in a narrower linewidth. But, as mentioned before, such cells give no EIT signal. Therefore, only CPT results are shown in Fig.\ \ref{cptbuffer}. The linewidth of the CPT resonance decreases from 15 kHz in a pure cell to 8 kHz in a buffer cell. As mentioned before, the absorption through such a cell off resonance is about 3\%. Hence the photodiode signal is adjusted to reflect this.

\subsection{Fluorescence detection}

We have verified that CPT resonances can be seen in the fluorescence signal from the cell, while EIT is not. But the detection (with a photomultiplier tube) is complicated by the fact that the cell is inside a multilayer magnetic shield. In addition, the powers in the beams have to be increased to get a good signal, which results in a large increase in the resonance linewidth. Hence the curve is not shown.

\subsection{Bright state}
Just like two phase-coherent beams can put the atom in a dark state, a similar arrangement can be used to create a bright state. The result is enhanced absorption at line center, exactly opposite to the reduced absorption seen in CPT. The first observation of this was in the closed $F_g=3 \rightarrow F_e = 4 $ transition in $^{85}\rm Rb $ \cite{LBA99}. The authors called the phenomenon electromagnetically induced absorption (EIA)---to highlight the fact that there was enhanced absorption on resonance. However, we feel that a more appropriate term would be CBS---standing for \textbf{c}oherent-trapping in a \textbf{b}right \textbf{s}tate---while the term EIA should be reserved for enhanced absorption of a weak probe laser in the presence of two or more strong control beams in multilevel systems, as demonstrated by us in Refs.\ \cite{CPN12,BHN15} and others in Refs.\ \cite{GWR04,BMW09}.

The conditions for observing CBS on a $ F_g \rightarrow F_e $ transition are:
\begin{enumerate}[(i)]
	\item It is a closed transition.
	\item $ F_e = F_g +1 $.
	\item $ F_g \neq 0 $, so that there are multiple magnetic sublevels in the ground state.  
\end{enumerate}
All these conditions are met for the $ 2 \rightarrow 3 $ transition in $ ^{87}\rm Rb $. A detailed study of CBS results will be presented in a future publication.

\section{Theoretical analysis}

The experimentally observed line shapes for both CPT and EIT can be obtained from a density matrix analysis of the three-level lambda system. In terms of the various density matrix elements, the time evolution is given by the following set of equations:

\begin{equation}
\begin{aligned}
	\dot{\rho}_{11} & =  -\Gamma_{11}\rho_{11}+{\Gamma_{21}}\rho_{22}+\frac{i}{2}\left(\Omega^{*}_{p}\rho_{12}-\Omega_{p}\rho_{21}\right) \\
	\dot{\rho}_{22} & =  -\Gamma_{22}\rho_{22}+\frac{i}{2}\left(\Omega_{p}\rho_{21}-\Omega^{*}_{p}\rho_{12}\right)\\
	&+\frac{i}{2}\left(\Omega_{c}\rho_{23}-\Omega^{*}_{c}\rho_{32}\right)  \\
	\dot{\rho}_{33} & =  -\Gamma_{33}\rho_{33}+{\Gamma_{23}}\rho_{22}+\frac{i}{2}\left(\Omega^{*}_{c}\rho_{32}-\Omega_{c}\rho_{23}\right) \\
	\dot{\rho}_{12} & =  \left(-\frac{\Gamma_{11}+\Gamma_{22}}{2}+{i}\Delta_{p}\right)\rho_{12}
	\\&+	\frac{i}{2}\Omega_{p}\left(\rho_{11}-\rho_{22}\right) 
	+ \frac{i}{2}\Omega_{c}\rho_{13}  \\
	\dot{\rho}_{13} & =  \left(-\frac{\Gamma_{11}+\Gamma_{33}}{2}+{i}(\Delta_{p}-\Delta_{c})\right)\rho_{13}\\
	& + \frac{i}{2}(\Omega^{*}_{c}\rho_{12}-\Omega_{p}\rho_{23})  \\
	\dot{\rho}_{23} & =  \left(-\frac{\Gamma_{22}+\Gamma_{33}}{2}-{i}\Delta_{c}\right)\rho_{23}\\
   & +\frac{i}{2}\Omega^{*}_{c}\left(\rho_{22}-\rho_{33}\right) - \frac{i}{2}\Omega^{*}_{p}\rho_{13} 
\label{densitymatrixeqns}
\end{aligned}
\end{equation}
There are only 6 equations for the 9 elements because the off-diagonal elements are complex conjugates of each other. The equations are solved in steady state, i.e.\ $\dot{\rho} = 0 $. Parameters corresponding to the experimental situation are: $ \Gamma_{11}/2\pi = 1 $ kHz, $ \Gamma_{22}/2\pi = 6$ MHz, $ \Gamma_{33}/2\pi = 1 $ kHz, $ \Gamma_{21}/2\pi = 3$ MHz, and $  \Gamma_{23}/2\pi = 3$ MHz.

For CPT, the quantity of interest is the upper state occupation, given by $ \rho_{22}$. The above equations are solved numerically, so that all orders of Rabi frequency are taken into account. The spectrum is Doppler averaged over the Maxwell-Boltzmann velocity distribution corresponding to room temperature vapor.

\begin{figure}
\centering{\resizebox{0.95\columnwidth}{!}{\includegraphics{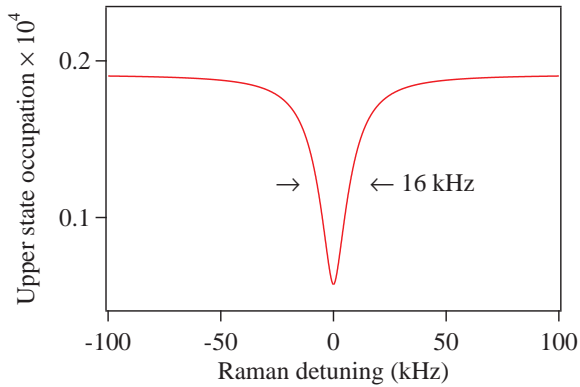}}}
\caption{(Color online) Calculated spectrum for the CPT resonance. The upper state occupation, given by $\rho_{22}$, is shown as a function of Raman detuning between the two beams. Both beams are taken to have equal powers of $0.04 \, \Gamma_{22}$. The occupancy reduces significantly when the dark state is formed.}
 \label{cpttheory}
\end{figure}

The experimental CPT spectrum in Fig.\ \ref{cptpure} is obtained with equal control and probe powers of 30 \textmu W, which corresponds to a maximum intensity at the center of the Gaussian beam(s) of 0.64 mW/cm$^2$. This is reduced by a factor of 10 so that the spectrum does not change significantly when the upper state linewidth $\Gamma_{22}$ is changed, which is required because the dark state does not involve the upper state. The other thing needed in the calculation is the decoherence rate between the lower levels. This is taken to be 1 kHz, adjusted so that the theoretical linewidth is comparable to the experimental one. This value is reasonable for the wall-collision rate in our cell. The resulting spectrum calculated with above parameters is shown in Fig.\ \ref{cpttheory}.

For EIT, the absorption of the probe beam is determined by the imaginary part of the coherence between levels $\ket{1}$ and $\ket{2}$, and is therefore given by $\Im \{\rho_{12}\Gamma_{22}/\Omega_{p}\}$. Since EIT is studied in the weak probe regime, the density matrix equations are solved to first order in the probe Rabi frequency. The steady state solution is given by: 
\begin{equation}
	\rho_{12} = \frac{{i} \Omega_{p}/2}{\displaystyle{\frac{\Gamma_{22}}{2}-{i}\Delta_{p}+\frac{{i}|\Omega_{c}|^{2}/4}{(\Delta_{p}-\Delta_{c})}}}
\label{solutionforrho12}
\end{equation}

\begin{figure}
\centering{\resizebox{0.95\columnwidth}{!}{\includegraphics{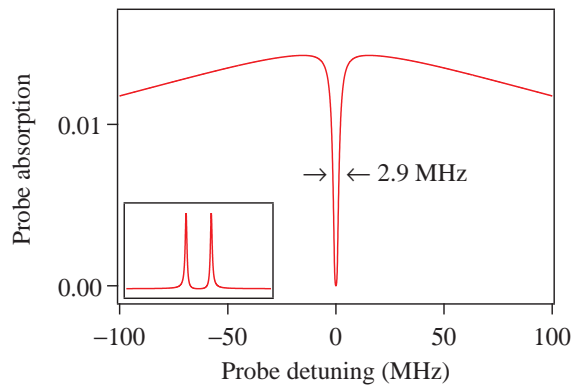}}}
\caption{(Color online) Calculated spectrum for the EIT resonance. Probe absorption, given by $ \Im \{ \rho_{12} \Gamma_{22} / \Omega_p \}$, is plotted as a function of probe detuning. The calculation is with $ \Omega_c = 5.7 \, \Gamma_{22} $ and $ \Omega_p = 0.05 \,
\Gamma_{22}$. The inset shows the result without Doppler averaging, which gives an Autler-Townes doublet separated by $ \Omega_c $.}
 \label{eittheory}
\end{figure}

For the experimental spectrum shown in Fig.\ \ref{eitpure}, the control power is 10 mW. This gives a maximum intensity at the center of the Gaussian beam of 212.3 mW/cm$^2$. Taking 50\% of this value as the average intensity over the entire beam, we get an intensity of 106.1 mW/cm$^2$. This corresponds to a Rabi frequency of $ \Omega_c= 5.7 \, \Gamma_{22} $. The resulting spectrum calculated with value of $ \Omega_c $ is shown in Fig.\ \ref{eittheory}. 

There are two things to note about the calculated spectrum: 
\begin{enumerate}[(i)]
	\item The Doppler averaging results in significant linewidth reduction, as demonstrated by us in Ref.\ \cite{IKN08}. In fact, for this value of Rabi frequency, the Autler-Townes doublet peaks are separated by $ 5.7 \, \Gamma_{22} $, which results in an EIT resonance that is neither subnatural nor Lorentzian. This is evident from the inset shown in the figure.
 
 	\item The observed spectrum is a convolution of the EIT spectrum with the linewidth of the probe laser. This means that 1 MHz should be added to the calculated linewidth, since this is not accounted for in the calculation. With this correction, the theoretical and experimental linewidths agree quite well.
\end{enumerate}

\section{Conclusions}

To conclude, we have highlighted the differences between the related phenomena of coherent population trapping, which requires two phase-coherent beams of roughly equal power, and electromagnetically induced transparency, which requires independent lasers with large power in the control beam and negligible power in the probe beam. 

The literature is full of cases where the term EIT is used for experiments that are of the CPT kind. Perhaps this is because of the pizazz associated with the phrase EIT. But an immediate clue as to the kind of experiment can be had from the number of lasers used---one laser implies CPT while two lasers imply EIT. The two phase-coherent beams from the same laser required for CPT are then produced by modulation, either acousto-optic or electro-optic. CPT experiments gain from the use of vapor cells filled with buffer gas because this increases the ground-state coherence time and hence narrows the linewidth of the resonance, whereas EIT cannot be seen in such a cell. Under the right conditions, the control laser in EIT leads to subnatural resonances for probe absorption. We have recently shown that the use of a control beam with a Laguerre-Gaussian profile---instead of the usual Gaussian---leads to a further narrowing with an unprecedented linewidth of $\Gamma/20$ being obtained \cite{CHN13}. 

\section*{Acknowledgments}
This work was supported by the Department of Science and Technology, India. S K acknowledges financial support from INSPIRE Fellowship, Department of Science and Technology, India. The authors thank S Raghuveer for help with the manuscript preparation, and E.\ Arimondo for helpful discussions and a critical reading of the manuscript.

\section*{Author contributions}
VN conceived the study and wrote the initial draft of the paper; SK and MPK did the experiments; SK and VB did the theoretical simulation; all authors reviewed the manuscript for intellectual content, and read and approved the final version.

\end{document}